\documentstyle[PASJadd,epsbox]{PASJ95}

\newcommand{\vol}{}
\newcommand{\no}{}
\newcommand{\lett}{}
\newcommand{\spage}{}
\newcommand{\rdate}{}
\newcommand{\adate}{}

\def\tj{$\theta_{\rm{jet}}$}
\def\bj{$\beta_{\rm{jet}}$}
\def\deg{$^{\circ}$}
\def\al{\alpha_{\rm vis}}

\begin{document}

\title{Sub-parsec-scale Acceleration of the Radio Jet
       in the Powerful Radio Galaxy NGC 6251}

\author{Hiroshi {\sc Sudou},$^1$ Yoshiaki {\sc Taniguchi},$^1$
        Youichi {\sc Ohyama},$^2$ Seiji {\sc Kameno},$^2$  \\
	Satoko {\sc  Sawada-Satoh},$^2$ Makoto {\sc Inoue},$^2$ 
        Osamu {\sc Kaburaki},$^1$  and Tetsuo {\sc Sasao}$^2$
	\\[12pt]
$^1${\it Astronomical Institute, Graduate School of Science,
       Tohoku University, Aoba, Sendai 980-8578}\\
{\it E-mail (HS): sudou@astr.tohoku.ac.jp}\\
$^2${\it National Astronomical Observatory of Japan, 2-21-1 Osawa,
       Mitaka, Tokyo 181-8588}}


\abst
{
In order to investigate the genesis of powerful radio jet, we have
mapped the central 10 pc region of the nearby radio galaxy NGC 6251
with a 0.2 pc resolution using Very Long Baseline 
Interferometer (VLBI) at two radio frequencies, 5 GHz
and 15 GHz, we have found the sub-parsec-scale counterjet
for the first time in this radio galaxy. This discovery allows us to
investigate the jet acceleration based on the relativistic beaming
model.
}

\kword{
galaxies: individual (NGC 6251) ---
galaxies: active ---
galaxies: jet ---
galaxies: nuclei ---
radio continuum: galaxies 
}

\maketitle
\thispagestyle{headings}

\section{Introduction}

The powerful radio jets are observed in approximately ten percent
of luminous AGN and their maximum lengths sometimes amount to
$\sim$ 1 Mpc (Bridle, Perley 1984).
Although global morphological properties give us very important
information, very inner regions in the radio jets also provide
hints for understanding the genesis of radio jets.
The capability of Very Long Baseline Interferometry (VLBI)
at radio frequency has been very useful in investigating
such inner regions of relatively nearby radio galaxies (Zensus 1997).
It has grown significantly since the
launch of the space radio telescope HALCA (Highly Advanced Laboratory
for Communications and Astronomy); a key element of the VLBI Space
Observatory Program (VSOP)
system (Hirabayashi et al. 1998; this volume). This has enabled
us to investigate 
radio galaxies at the sub-milli arcsecond angular resolution, although the
resolution depends on the radio frequency.

Since it is known that powerful radio jets are often emanated from the
central engine of AGN at a relativistic speed, the inner region of
radio jets is just a very exciting place. Since a pair of radio lobes
are generally observed in the radio galaxies, the jet is believed to
be intrinsically ejected in two opposite directions. However,
although the approaching jet is enhanced by the Doppler beaming,
the receding one (the counterjet) is significantly extinguished if the jet
velocity is highly relativistic and is observed from a smaller
viewing angle toward the jet. Indeed, in many radio galaxies,
the counterjet is hardly seen in the vicinity of the radio core.
Since a typical distance of the radio jets resolved by VLBI 
is $\sim $ 1 pc inner region, this inner region is called as
pc-scale jets (Zensus 1997).

NGC 6251 is one of apparently brightest powerful radio galaxies
in the nearby universe and thus has been investigated extensively
using VLBI as well as VLA (Very Large Array)
(Waggett et al. 1977; Cohen, Readhead 1979;
Perley et al. 1984; Jones et al. 1986; Jones, Wehrle 1994).
So far these measurements failed to detect the pc-scale
counterjet. If we could detect it, it will be possible to give
many important observational constraints on the jet geometry
and then the physical process of jet acceleration.
If the jet acceleration would occur at a sub-pc region,
we could detect the counterjet because the
fading due to the Doppler beaming is still insignificant.
In order to find evidence for such sub-pc-scale counterjet in NGC 6251,
we have performed new high-resolution VLBI observations using HALCA.
We use a distance to NGC 6251, 94.4 Mpc, which is determined with
the use of a recession velocity to the galactic standard of rest,
$V_{\rm GSR}$ = 7079 km s$^{-1}$ (de Vaucouleurs et al. 1991)
and a Hubble constant  $H_0$ = 75 km s$^{-1}$ Mpc$^{-1}$.
Note that 1 mas (milli arcsecond) corresponds to 0.48 pc at this distance.

\section{Observations}

NGC 6251 was observed at 5 GHz using VSOP on 30 April 1998
(10.1 h on-source) and at 15 GHz using VLBA (Very Long Baseline Array)
on 2 June 1998
(9.3 h on-source). The VSOP system used in our
observation consists of HALCA, VLBA, and Effelsberg 100-m radio
telescope near Bonn. Details of the observations are summarized in table
1.
Since quality of the data taken at the Mauna Kea station
of VLBA appeared to be
poorer than those taken at the other stations, we flagged out
the visibilities with the Mauna Kea station.
The band width of the VSOP receivers is 32 MHz, divided into two
IF bands, each of which has 128 channels at 5 GHz.
The bandwidth of the VLBA receivers is also 32 MHz but
divided into four IF bands each of which has 16 channels at 15 GHz.
We used the source 1803+784 as a calibrator at both frequencies.
Correlation processes have been carried out
using the VLBA correlater at Soccoro in New Mexico.
The integration time at the correlater was 3 s.
We used the NRAO AIPS package for fringe fitting and a-priori
amplitude calibration.
Image restoration via self-calibrations
were made using DIFMAP. The resulting
visibilities were coherently averaged for 30 s.
The synthesized beam size of the two images are 0.73 $\times$ 0.34 mas in
PA = $-9.2^\circ$ at 5 GHz and 0.46 $\times$ 0.38 mas
in PA = $-13.4^\circ$ at 15 GHz with uniform weighting.
In order to perform
beam-size-matched comparison between 5 GHz and 15 GHz, we restored the
two images with a same spatial resolution of 0.50 $\times$ 0.50 mas.
The {\it uv} coverages of the two observations are shown in figure 1.


\begin{table*}
\small
\begin{center}
Table~1.\hspace{4pt}Observations.\\ 
\end{center}
\vspace{6pt}
\begin{tabular*}{\textwidth}{@{\hspace{\tabcolsep}
\extracolsep{\fill}}cccccc}
\hline\hline\\[-6pt]
 Observations & $\nu$ & Stations & Peak Int. & RMS noise & Dynamic \\
              & (GHz) &          & (Jy/beam) & (mJy/beam)& Range\\
[4pt]\hline\\[-6pt]
  1998 Apr. 30& 5&  VLBA, HALCA$^*$, EB$^{\dag}$&  0.13& 0.82& 160 \\
  1998 Jun. 02& 15& VLBA&                     0.34& 0.24& 1400 \\[4pt]
\hline
\end{tabular*}

\vspace{1pt}

\noindent
$^*$Highly Advanced Laboratory of Communication and Astronomy\\
$^{\dag}$Effelsberg 100-m telescope
\end{table*}

\begin{figure}
 \begin{center}
\epsfile{file=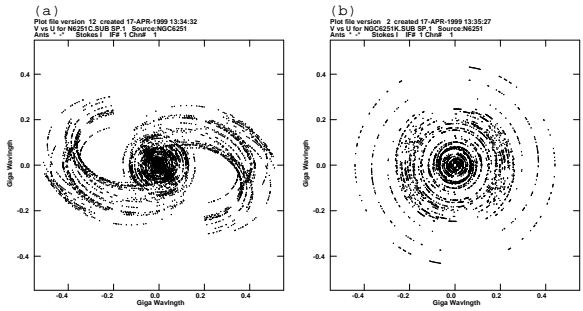,scale=0.4}
\caption{The {\it uv} coverages at 5 GHz (a) and at 15
GHz (b).}
  \end{center}
\end{figure}

\section{Results}

Our final maps at 5 GHz and 15 GHz are shown in figure 2a and 2b,
respectively. The dynamic ranges are 160 for the
5 GHz map and 1400 for the 15 GHz map. In both maps, several knots
are found
within 10 mas (4.8 pc) from the each brightest peak.
Analyzing the wiggling patterns in both frequencies, we obtain
an opening angle of the jet, $\phi \simeq 0.50^\circ \pm 0.29^\circ$.
It is remarkable that a counterjet-like feature can be seen in both
maps although its spatial extension is only $\approx$ 1 mas.
The most prominent knot in the 5 GHz map is seen at $\approx$ 2.0 mas
(0.96 pc) away from the 5 GHz brightest peak. Its intensity is $\simeq$ 26\%
of the 5 GHz core. This knot is found for the first time.
It is also seen in the 15 GHz map. However, its angular
distance from the 15 GHz brightest peak is $\approx$ 2.3 mas which is
larger by 0.3 mas (0.14 pc) than that in the 5 GHz map.

Since any VLBI images have no information on the absolute position
because of the self-calibration procedure,
this difference makes it difficult to register the two images
precisely. Since it is often observed that radio cores are optically
thick even at those frequencies because of higher plasma densities,
it is  unlikely that the observed brightest peak positions always
correspond to the real core position; here we assume that the very
center of the active nucleus, i.e., the central engine is enshrouded by
the VLBI core. 
On the other hand, knots are
thought to be optically thin and thus their brightness distributions
show more realistic distributions of plasma.
Therefore, it is better to register the two images using
the spatial and morphological information of knots between the two maps.

Here we apply two new methods for the registration.
The first one is to analyze the correlation function between
intensity profiles of the radio jet at the two frequencies.
The correlation function exhibits the maximum correlation
coefficient $C_{\rm int} \simeq 0.98$ at an offset of $\Delta_{\rm{int}}
\simeq 0.30 \pm$ 0.0025 mas, indicating that the peak at 5 GHz is shifted
by $\Delta_{\rm{int}}$ to the jet direction from that at 15 GHz (figure 3a).
The second method is to analyze the correlation  function
between wiggle patterns of the radio jet at the two frequencies.
This analysis gives an offset of $\Delta_{\rm wig} \simeq 0.10$
$\pm$ 0.24 mas  with the maximum correlation coefficient
of $C_{\rm wig} \simeq$ 0.34 (figure 3b). Since an offset of
$\Delta_{\rm{int}}$ is in good agreement with  that of $\Delta_{\rm
wig}$ within their errors,
we adopt a $\Delta = 0.3$ mas as the best offset.
Using this offset, we register the two images and make a
spectral index image of the radio jet of NGC 6251 shown in figure 2c.
Here the spectral index $\alpha$ is defined as $S_\nu \propto
\nu^{\alpha}$.
It is shown that the core region has an inverted spectrum
with $\alpha \simeq 1.2$ and thus it appears optically thick.
On the other hand, the spectral index of the jet ranges 
$-$1 and $-$0.5 and thus the jet is optically thin. 
The counterjet-like feature has a spectral index of $\simeq -0.5$
and thus appears to be optically thin. This ensures that it 
is the counterjet.  Its intensity level
is $\sim 10 \sigma_{\rm rms}$ at 5 GHz and $\sim 40 \sigma_{\rm rms}$ at
15 GHz, where $\sigma_{\rm rms}$ is the one sigma root-mean-square noise 
in each map. 
Therefore, we conclude that this is the real counterjet.
Curiously, the morphology of the counterjet is
slightly different between the two images, i.e., two ridge lines can be
seen only at 15 GHz. Although it is difficult to understand 
such a morphological difference, we confirmed that one of ridge lines at
15 GHz corresponds to the counterjet  at 5 GHz.

\begin{figure}
 \begin{center}
\epsfile{file=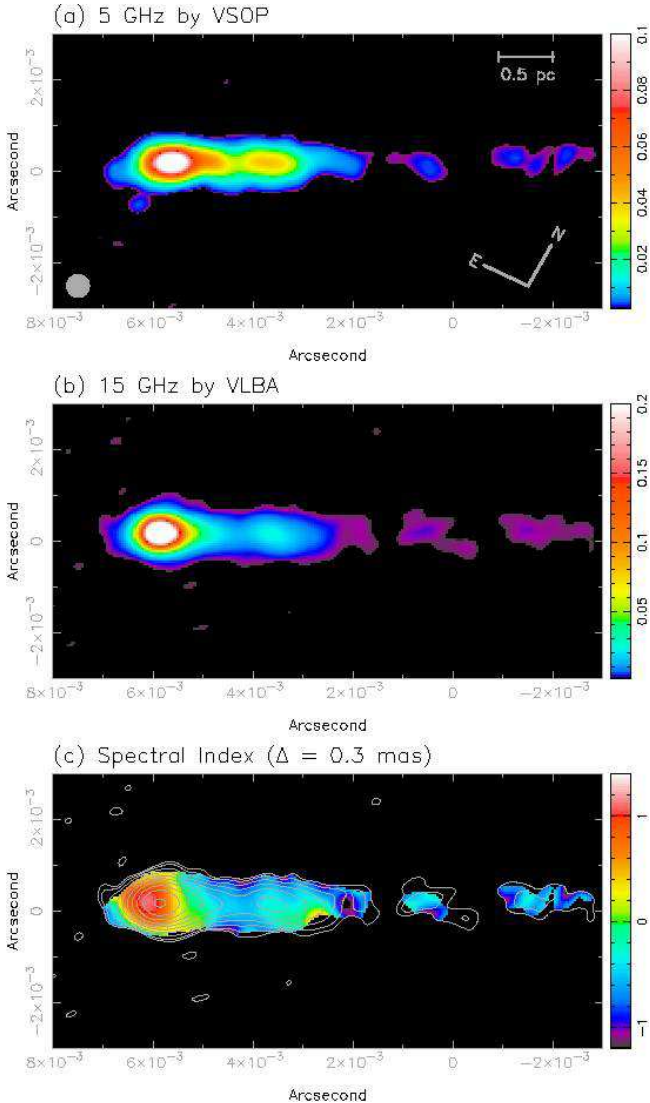,scale=0.55}
\caption{The images of NGC 6251 at 5 GHz (a) and at 15 GHz (b).
The spectral index map together with the 15 GHz contour image is also
shown in (c). All the  
images rotated clockwise on the sky by $28.4^\circ$. }
  \end{center}
\end{figure}

\begin{figure}
 \begin{center}
\epsfile{file=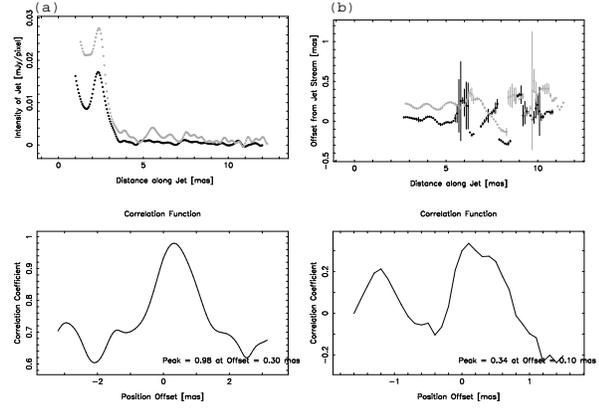,scale=0.4}
\caption{{\it Top:} Intensity profiles (a) and wiggling patterns (b)
at 5 GHz (grey circle) and 15 GHz  
  (filled circle). They  are already
  registered.  {\it Bottom}: Correlation coefficient as a function of
  offset.  }
  \end{center}
\end{figure}

\section{Discussion}

\subsection{The Absorption Model at the Core}

The inverted spectrum in the core implies that there is a strong
absorption at 5 GHz. This may explain why
the brightest position at 5 GHz does not coincide with that at 15 GHz.
In general, such absorption in the radio jet is caused either
by free-free absorption or by synchrotron self-absorption. Since it
seems quite
likely that a plasma halo (or ring) surrounds the central engine, the
free-free absorption seems to be a dominant absorption
mechanism, i.e., the flux density at 5 GHz is more significantly
absorbed by the 
free-free absorption than the 15 GHz brightest peak flux.

Free-free
absorption models give a  
relation $S_\nu = S_0
\nu^{\alpha_0} \exp [- \tau_0 \nu^{-2.1}]$. Accordingly,
we obtain,

\begin{equation}
 \tau_{\rm obs} = \frac{\alpha^{15}_5 - \alpha_0}{0.0279}
  \left({\frac {\nu}{{\rm GHz}}}\right)^{-2.1},
\end{equation}
where $\alpha^{15}_5$ is the spectral index between 5 and 15 GHz and
$\alpha_0$ is the intrinsic spectral index of the jets.
Using $\alpha^{15}_5 = 1.2$,  $\alpha_0 = -0.5$, and $\nu = 5
$ GHz,  we obtain $\tau_{\rm obs} = 2.4$.

A question arises as how the absorbing material is distributed around
the central engine at a sub-pc-scale. Since we do not know the true
situation, we will investigate the following three absorption models:
(1) a spherical absorber model, (2) an optically-thick disk model, and
(3) an optically-thin disk model.

\subsubsection{The spherical absorber model}

First, we investigate a spherical absorber model because this model is
the simplest one. 
We assume that the core is surrounded by a plasma sphere 
with a radius of $a$ and the radio emission from the inner part of
the jets (i.e., both the jet and the counterjet) suffer from 
the free-free absorption. We show this model schematically
in figure 4. In this model, we adopt the 15 GHz brightest peak 
as the core in which the central engine resides.  
The approaching jet escapes from the plasma sphere at a projected 
distance of $x = a \times \sin \theta_{\rm jet}$ and thus the 
optical depth becomes to be small (i.e., $\tau_{\rm obs} < 1$) here.
On the other hand, although the counterjet escapes the sphere
at  $x = - a \times \sin \theta_{\rm jet}$, it suffers the effect of
free-free absorption until $x = -a$. It is also noted that
the path length is longest at $x = - a \times \sin \theta_{\rm jet}$ .
As shown in figure 4, we define the following projected distances;
$X_{\rm jet} = X_{\rm peak} =  a \times \sin \theta_{\rm jet}$
and $X_{\rm cjet} = a$.
Then we are able to estimate \tj,

\begin{equation}
 \theta_{\rm jet} = \sin^{-1}
  \left({\frac {X_{\rm jet}}{X_{\rm cjet}}}\right)
\end{equation}

In figure 5, we show the observed spectral index variation
along the radio jet. 
The opacity is assumed to be proportional
to the path length through the plasma sphere.
The region with an inverted spectrum is considered to be optically thick
for the free-free absorption ($\tau_{\rm obs} > 1$) while that
with a normal spectrum is optically thin. We
adopted that $\tau_{\rm obs}$ = 1 when the jets
escape from  the sphere.  From equation (1), note that $\tau_{\rm obs}$
= 1 at 5 GHz  corresponds to  
$\alpha_5^{15} = 0.2$ and that $\tau_{\rm obs}$ at 15 GHz
is always smaller than unity in our observations. 
Therefore, from figure 5, we estimate $X_{\rm peak} \approx$ 0.20 mas,
$X_{\rm jet} \approx$ 0.25 mas, and $X_{\rm cjet} \approx$ 0.75 mas.
Then we obtain $\theta_{\rm jet} \simeq 19^\circ$.
In the previous VLBI measurements of the pc-scale jet of NGC 6251,
Jones (1986) obtained $\theta_{\rm jet} \leq 45^\circ$
using the upper limit of the counterjet flux.
Our estimate appears consistent with their result.
Although our simple model suggests $X_{\rm jet} = X_{\rm peak}$,
the observation results in $X_{\rm jet} / X_{\rm peak} \approx 1.3 \ne 1$.
This may be caused by plasma density fluctuation in the actual radio core
of NGC 6251. If we adopt the observed value of $X_{\rm jet}$ instead of
$X_{\rm peak}$, we obtain $\theta_{\rm jet} \simeq 15^\circ$. 

\begin{figure}
 \begin{center}
\epsfile{file=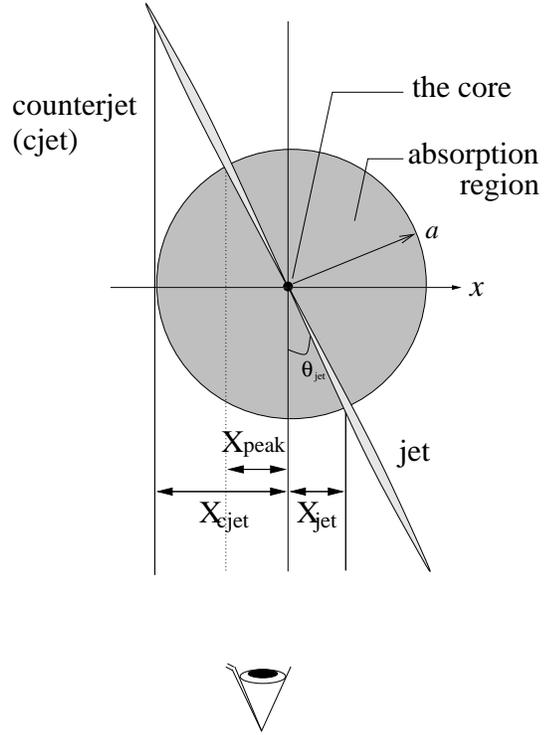,scale=0.4}
\caption{A schematic illustration of the spherical absorber model in NGC 
6251.}
  \end{center}
\end{figure}

\begin{figure}
 \begin{center}
\epsfile{file=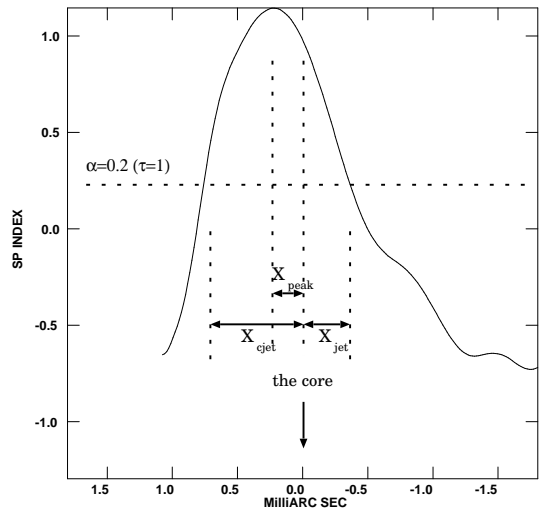,scale=0.65}
\caption{Spectral index variation along the jet near the core.}
  \end{center}
\end{figure}

\subsubsection{The optically-thick disk model}

Although the spherical absorber model is useful for estimating the jet
viewing angle,  it is too simple to determine the 
detailed environment of the radio core. 
Recently, it has been reported that a very
narrow gap near the core of the radio jet in NGC 4261 is attributed to
the absorption by 
a geometrically thin, nearly edge-on accretion disk (Jones et al.
2000). 
In this case, the central engine may be powered by gas accretion at
a sub-Eddington accretion rate. It is expected that  an
accretion structure consists of an optically-thin, advection-dominated
accretion flow (ADAF) in the central region and a geometrically thin
accretion disk surrounding it (e.g., Narayan et al.
1998). This idea can be applicable to the case of NGC 6251.

We consider that the central black hole is surrounded by 
such a tenuous plasma or coronal region which is related to ADAF, and that
a relatively dense outer disk 
causes a strong free-free absorption. In this model, the strongest
absorption should be observed at the inner edge of the outer
disk. Therefore, we identify the location of the maximum absorption at 5
GHz with the inner edge. If the jet viewing angle
$\theta_{\rm jet} = 30^\circ$ (see Sudou, Taniguchi 2000), then we
obtain the linear distance of the inner edge, 
$r_{\rm peak} \equiv X_{\rm peak}/\cos \theta \approx  3.3 \times
10^{17}$ {\rm cm}. 
This  model is shown schematically in figure 6.

According to the  usual understanding of  accretion disks (Narayan et
al. 1998), we regard the outer disk as the optically-thick disk
(Shakura, Sunyaev 1973). Since the density and the
temperature are low in this outer disk, the free-free absorption may be
dominant over the  electron scattering. 
Following Shakura and Sunyaev (1973),
the half thickness of the disk $H$,  the particle density $n$, and  the
electron temperature $T$ are given by,

\begin{eqnarray}
 H &=& 4.0 \times 10^{10} \al^{-1/10} \dot{m}^{3/20}
  m^{9/10} {\hat r}^{9/8}~{\rm cm},  \\
 n &=& 2.1 \times 10^{21} \al^{-7/10} \dot{m}^{11/20}
  m^{-7/10} {\hat r}^{-15/8}
  ~{\rm cm}^{-3}, \\~{\rm and }\\
 T &=& 9.8 \times 10^6 \al^{-1/5} \dot{m}^{3/10}
  m^{-1/5} {\hat r}^{-3/4} ~{\rm K},
\end{eqnarray}
respectively, where $\al$ is the viscosity parameter, and $m$ is the black
hole mass,  ${\dot m}$ is the mass accretion rate, and ${\hat r}$ is the
radial distance. The latter three parameters are scaled 
for a typical AGN as $m \equiv M_{\rm BH}/10^8 M_{\odot}$, 
${\dot m} \equiv {\dot M}/{\dot M}_{\rm crit} ~({\dot M}_{\rm crit}
\equiv L_{\rm Edd}/0.1 c^2 \sim 1.5 \times 10^{26} m 
~{\rm g~s^{-1}})
$,  and ${\hat r} \equiv r/r_{\rm g}
~(r_{\rm g} \equiv 2GM/c^2 \sim 3.0 \times 10^{13} m ~{\rm cm})$.
For NGC 6251,  we adopt $m = 8$ (Ferrarese, Ford 1999), ${\dot m} =
0.001$ (Birkinshaw, Worral 1994), 
and  ${\hat r} = {\hat r}_{\rm peak} = 1.4 \times 10^3$. 
Assuming $\al = 0.1$, we obtain, 
 $ H = 4.0 \times 10^{14}$~{\rm cm},  
 $ n = 7.0 \times 10^{13}$ ~{\rm cm}$^{-3}$, and
 $ T = 5.5 \times 10^3$  ~{\rm K}.
Although $H$ and  $T$ are reasonable compared with
observations,
$n$ is too large as to given a very high opacity of $\tau \sim 10^{16}$
at 5 GHz.
Therefore, this model fails to explain the observational properties of
NGC 6251.

\begin{figure}
 \begin{center}
\epsfile{file=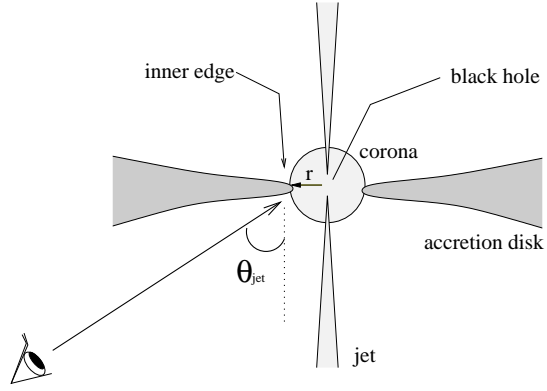,scale=0.45}
\caption{A schematic illustration of the absorption disk model in NGC 
6251.}
  \end{center}
\end{figure}

\subsubsection{The optically-thin disk model}

Next, we attempt to explain $\tau_{\rm obs}$ adopting an optically-thin
disk model in which the viscous heating is balanced by the 
free-free cooling.
Adopting the cooling rate of $1.2 \times 10^{21} m_{\rm p}^2 n^2 T^{1/2} H$
erg s$^{-1}$ cm$^{-2}$ (Kato et al. 1998), 
we obtain,

\begin{eqnarray}
 H &=& 5.8 \times 10^{12} \al^{-1/2} \dot{m}^{1/4}  
  m^{1} {\hat r}^{9/8}~{\rm cm},  \\
 n &=& 5.3 \times 10^{13} \al^{1/2} \dot{m}^{1/4}
  m^{-1} {\hat r}^{-15/8} ~{\rm cm}^{-3}, ~{\rm and }\\
 T &=& 1.0 \times 10^{12} \al^{-1} \dot{m}^{1/2}
  m^{0} {\hat r}^{-3/4} ~{\rm K}.
\end{eqnarray}
Adopting the same parameter values as those in section 4.1.2, we obtain  
 $H = 9.0 \times 10^{16}$ ~{\rm cm},  
 $n = 4.7 \times 10^{5}$ ~{\rm cm}$^{-3}$, and 
 $T = 1.4 \times 10^9$  ~{\rm K} for NGC 6251. 
The derived temperature of $T \sim 10^9$ K is much higher than
that of $\sim 10^4$ K assumed in the model of Jones et al. (2000). 
In view of the resulting high electron temperature, the cooling in this
situation may be dominated by the inverse 
Compton scattering rather than by free-free radiation.  
The model suitable for such a condition may be that of Shapiro et
al. (1976) which adopts that the radiation cooling is 
dominated by inverse Compton scattering. In fact, the
expected free-free optical depth in our model is only $\sim
10^{-6}$. Therefore, we conclude 
that this model is unsuitable for the absorption mechanism of NGC
6251.   


\subsubsection{Summary of the absorption models}

The simplest spherical absorption model provides the plausible
explanation for the observed absorption properties. On the other hand,
the optically-thick and optically-thin disk models fail to explain the
observations, although the presence of the disk structure would be
naturally expected around the central engine. It will be necessary to
seek for  other types of outer accretion-disk models which are
consistent with the observations.

\subsection{Velocity of the Jet}

Adopting the so-called Doppler beaming model (e.g., Ghisellini et al. 1993), 
we estimate the jet velocity using the observed jet-counterjet
intensity ratio; i.e., if
the jet of NGC 6251 is intrinsically two-directional symmetric,
the observed intensity asymmetry between the jet and counterjet
can be attributed to the Doppler beaming.
The jet-counterjet intensity ratio $R$ is related to
both the intrinsic jet velocity $v_{\rm jet}$ and the viewing angle \tj 
toward the jet by the formula

\begin{equation}
R = \left[\frac {1 + \beta_{\rm{jet}} \cos \theta_{\rm{jet}}}
   {1 - \beta_{\rm{jet}} \cos \theta_{\rm{jet}}}\right]^{3 - \alpha_0},
\end{equation}
where \bj = $v_{\rm jet} / c$.
Using the intensity profiles of both 5 GHz and 15 GHz along the jet
direction, we have measured $R$ and plot them in figure 7. Note
that the 
core component ($|x| \leq$ 0.5 mas) is not used  in this analysis.
Although the estimated value of $R$ at 1.05 mas from the core is
the same at 
both frequencies, that within 0.8 mas from the core is higher at 5 GHz
than  at 15 GHz. 
Since this is probably due to severer absorption at 5 GHz 
than at 15 GHz, the value of $R$ at 15 GHz appears to be more reliable.
It is shown that $R$ increases with
increasing projected distance from the core. 
If this increase is caused mainly by the Doppler beaming, it is suggested
that \bj ~ and/or \tj ~ increase with increasing distance.
Since it is unlikely that \tj ~ varies significantly at such a small
scale,
it seems reasonable to conclude that
the increase of $R$ is attributed to the increase in \bj. This can be
interpreted as possible evidence for the 
acceleration at the sub-pc-scale radio jet. 
Using equation (9) and assuming \tj = 30 \deg, we show that the jet
is accelerated from \bj ~ 
$\approx 0.13$ at 0.55 mas to \bj ~ $\approx 0.42$ at 1.05 mas. The
results are  summarized in table 2.
These angular distances correspond to the linear distances of
$\approx$ 0.30 pc and $\approx$ 0.58 pc, respectively.

If the jet acceleration is effectively made at sub-pc-scale region,
the jet velocity becomes relativistic at radial distance of $\sim$ 1
pc. If the sub-pc-scale acceleration is a general property in most radio 
galaxies, it seems difficult to detect pc-scale counterjets. It will be
very important to observe sub-pc-scale jets at higher frequency because
the effect of absorption is negligibly small. 

Since the kinetic energy of a particle with a mass of $m$ 
at a relativistic velocity of $v$ is given as  $E = m \gamma v^2
= m \gamma \beta^2 c^2$ where $\gamma = (1 - \beta^2)^{-1/2}$, 
the kinetic energy gain from $\beta = 0.1$
to $\beta = 0.4$ per unit mass is $\Delta E/m \sim 3 \times 10^{16}$
J kg$^{-1}$. This corresponds to the kinetic energy of $\sim 
 200$ keV for an electron and to that of $\sim$ 300 MeV for a proton.
Therefore, it is suggested that the jet acceleration region
found in NGC 6251 is related to the $\gamma$-ray emitting region.
This may explain why $\gamma$-ray loud AGNs tend to be associated with
radio-loud AGNs if such acceleration occurs commonly in radio
galaxies (Shrader et al. 1995).


\begin{table*}
\small
Table~2.\hspace{4pt}The jet/counterjet intensity ratio.\\
\vspace{6pt}
\begin{tabular*}{\columnwidth}{@{\hspace{\tabcolsep}
\extracolsep{\fill}}ccccc}
\hline\hline\\[-6pt]
Distance (mas) & $R_5^*$ & $R_{15}^{\dag}$ & $\beta_{\rm jet}^{\ddag}$\\
[4pt]\hline\\[-6pt]
 0.55 \dotfill &  6.8$\pm$0.4     & 2.2$\pm$0.015 & 0.13$\pm$0.0014 \\
 0.80 \dotfill &  9.0$\pm$1.1 & 6.3$\pm0.23$  & 0.30$\pm$0.0053 \\
 1.05 \dotfill &  13.5$\pm$3.0& 13.5$\pm$1.6  & 0.42$\pm$0.017 
\\[4pt]
\hline
\end{tabular*}

\vspace{1pt}

\noindent
 $^*$$R$ at 5 GHz\\
 $^{\dag}$$R$ at 15 GHz\\
 $^{\ddag}$Assuming $\theta_{\rm{jet}}$ = 30$^\circ$

\end{table*}

\begin{figure}
 \begin{center}
\epsfile{file=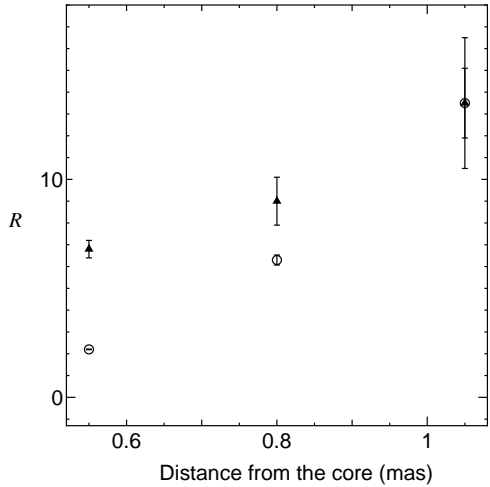,scale=0.4}
\caption{The jet-counterjet ratio $R$ as a function of the projected
distance from the core. The filled triangle and open circle indicate
obtained $R$ at 5 GHz and 15 GHz, respectively.}
  \end{center}
\end{figure}

\section{Conclusions}

We discussed high-resolution images of NGC 6251 at 5 GHz and 15
GHz. From the analysises of the intensity profile and the wiggling
pattern, we register the two images and  obtain the spectral index
image. It  shows  the counterjet which has an optically-thin  
spectrum. In order to explain the position difference between the core
and the absorption peak, we adopt the simple  
spherical absorption model, and show the possibility that the jet viewing
angle can be obtained from only the spectral index distribution. 
Although we also attempt to adopt the free-free absorption by an
accretion disk model, both optically thick and thin disk models 
based on the  $\alpha$-viscosity cannot explain the observed opacity. 

The jet-counterjet intensity ratio analysis shows 
that the radio jet of NGC 6251 is accelerating {}from 0.13 $c$ at the
linear radial distance of $r \approx$ 0.30 pc to 0.42 $c$  at $r
\approx$ 0.58 pc.  This provides the first direct evidence for the
sub-pc-scale acceleration of the radio jet in NGC 6251.

\vspace{1pc}\par
We would like to thank to staff of the VLBA and the VSOP for their
kind help 
of the observations. 
We also wish to thank D. Jones and H. Hirabayashi for their useful
comments and  encouragement. 
This work was financially supported in part by Grant-in-Aids for the
Scientific
Research (Nos. 10044052, and 10304013) of the Japanese Ministry of
Education, Culture, Sports, and Science. HS was supported by the
Grant-in-Aid for JSPS Fellows by Ministry of
Education, Culture, Sports, and Science. 


\section*{References}
\small
\re
Birkinshaw M., Worrall D.M. 1993, ApJ 412, 568
\re
Bridle A.H.,  Perley R.A. 1984,
              ARA\&A  22, 319
\re
Cohen M.H.,  Readhead A.C.S. 1979, ApJL
              233, L101
\re
de Vaucouleurs G., de Vaucouleurs A., Corwin, C.Jr., Buta R.J.,
              Paturel G., Fouq$\acute{\rm u}$e P. 1991, in Third
	      Reference Cataloue  of Bright Galaxies, (Springer-Verlag)
\re
Ferrarese L., Ford H.C. 1999, ApJ 515, 583
\re
Ghisellini G., Padovani P., Celotti A., Maraschi
              L. 1993, ApJ 407, 65
\re
Hirabayashi H., Hirosawa H., Kobayashi H., Murata Y.,
              Edwards P.G., Fomalont E.B., Fujisawa K., Ichikawa
              T. et al. 1998, Science 281, 1825
\re
Jones D.L., Unwin S.C., Readhead A.C.S., Sargent
              W.L.W., Seielstad G.A., Simon R.S., Walker R.C.,
              Benson J.M. et al. 1986, ApJ 305, 684
\re
Jones D.L. 1986, ApJL 309, L5
\re
Jones D.L.,  Wehrle A.E. 1994, ApJ 427, 221
\re
Jones D.L., Wehrle A.E., Meier D.L.,  Piner G.B.
              2000, ApJ 534, 165
\re
Kato S., Fukue J., Mineshige S. 1998, in Black-Hole Accretion Disks
              (Kyoto Univ. Press) p248
\re
Narayan R., Mahadevan R.,  Quataert E. 1998, in
              Theory of Black Hole Accretion Disks, ed M.A. Abramowicz,
              G. Biornsson, J.E. Pringle (Cambridge
              Univ. Press) p148
\re
Perley R.A., Bridle A.H.,  Willis A.G. 1984, 
              ApJS 54, 291
\re
Shakura N.I.,  Sunyaev R.A. 1973, A\&A 24, 337
\re
Shapiro S.L., Lightman A.P.,  Eardley D.M. 1976,
              ApJ 204, 187
\re
Shrader C.R., Gehrels, N. 1995, PASP 107, 606
\re
Sudou H., Taniguchi Y. 2000, AJ in press
\re
Urry C.M.,  Padovani P. 1995, PASP 107, 803
\re
Waggett P.G., Warner P.J.,  Baldwin J.E. 1977,
              MNRAS 181, 465
\re
Zensus J.A. 1997, ARA\&A 35, 607

\end{document}